\documentclass[11pt,letterpaper]{article}
\usepackage{float}
\newcommand {\be}{\begin{equation}}
\newcommand {\ee}{\end{equation}}
\newcommand {\bea}{\begin{eqnarray}}
\newcommand {\eea}{\end{eqnarray}}

\usepackage[dvips]{graphicx}
\graphicspath{{../eps/}}
\DeclareGraphicsExtensions{.eps}
\usepackage{epsfig,color}
\usepackage{booktabs}
\usepackage{epstopdf}
\usepackage{amssymb}
\usepackage{mathtools,cite}
\usepackage{xspace} 
\usepackage{algorithmic}

\usepackage{array}

\usepackage{mdwmath}
\usepackage{mdwtab}
\usepackage{bm}
\usepackage{mathrsfs}
\usepackage{epsfig}
\usepackage{subfigure}
\usepackage{algorithmic}
\usepackage{amsmath}
\usepackage{mathrsfs}
\usepackage{cleveref}  
\usepackage[section]{placeins}
\usepackage{geometry}
\newcommand{\R}[1]{\textcolor{black}{#1}}

\begin{document}
\begin{centering}
{\Large \bf
Storage capacity of networks with discrete synapses and sparsely encoded memories}\\
\vspace{0.5cm}
Yu Feng$^1$ and Nicolas Brunel$^{1,2}$\\
\vspace{0.25cm}
$^1$ Department of Physics, Duke University, Durham, NC 27710\\
$^2$ Department of Neurobiology, Duke University, Durham, NC 27710\\
\end{centering}

\begin{center}
{\bf Abstract}
\end{center}
Attractor neural networks (ANNs) are one of the leading theoretical frameworks for the formation and retrieval of memories in networks of biological neurons. 
In this framework,  a pattern imposed by external inputs to the network is said to be learned when this pattern becomes a fixed point attractor of the network dynamics. The storage capacity is the maximum number of patterns that can be learned by the network. In this paper, we study the storage capacity of fully-connected and sparsely-connected networks with a binarized Hebbian rule, for arbitrary coding levels.  Our results show that a network with discrete synapses has a similar storage capacity as the model with continuous synapses, and that this capacity tends asymptotically towards the optimal capacity, in the space of all possible binary connectivity matrices, in the sparse coding limit. We also derive finite coding level corrections for the asymptotic solution in the sparse coding limit. The result indicates the capacity of networks with Hebbian learning rules converges to the optimal capacity extremely slowly when the coding level becomes small. Our results also show that in networks with sparse binary connectivity matrices, the information capacity per synapse is larger than in the fully connected case, and thus such networks store information more efficiently. 

\section{Introduction}

It is widely believed that memories are stored in the brain through synaptic modifications in an activity-dependent way. This idea has been implemented in attractor neural network models, where the connectivity strength between neurons is determined by Hebbian synaptic plasticity  rules \cite{Hopfield, Tsodyks}. In this framework, a pattern is said to be learned if it becomes a fixed point attractor of the network dynamics. An extensively studied question is, how many patterns can be stored in such networks? Classical studies of memory modeling synapses as continuous variables in networks of binary neurons have shown that such networks can store a number of uncorrelated random patterns $p$ that scales linearly that network size, $p_{max}=\alpha_c N$ where $\alpha_c$ is of order 1 in the large $N$ limit \cite{Amit, Tsodyks, Gardner}. However, there is evidence suggesting that synapses in brain structures involved in memory, such as the hippocampus and neocortex, are more digital than analog \cite{Petersen, Connor, Dork,Hruska}. 

A number of studies have addressed the question of the storage capacity of networks with discrete synapses. Krauth and M\'ezard showed that networks can potentially have a large capacity when all synapses are required to be binary \cite{Krauth}, using Gardner's approach \cite{Gardner}, with an upper bound for capacity $\alpha_{cmax}=0.83$ discrete synapses, instead of $\alpha_{cmax}=2$ for continuous synapses. Sompolinsky studied the storage capacity of a network with a specific binarized Hebbian rule \cite{Sompolinsky_1,Sompolinsky_2}, and showed its capacity is remarkably close to the capacity of the Hopfield network \cite{Hopfield}, whose synapses are continuous variables ($\alpha_{c}=0.10$ instead of 0.14). However, these authors only studied the unbiased case, in which the coding level $f$ (i.e. fraction of active neurons in a pattern) is 0.5, while neuronal activity in areas involved in memory is typically very sparse (e.g.~\cite{Lee})- for instance, the coding level in the human medial temporal lobe has been estimated to be around 1\% \cite{Waydo}. The upper bound for capacity in networks with arbitrary coding levels and discrete synapses was computed by Gutfreund and Stein \cite{Gutfreund}. The capacity in networks with Hebbian plasticity and binary synapses has only been computed in the unbiased case, and the capacity for arbitrary coding levels remains an open question. In this paper, we generalized Sompolinsky$'$s calculation on the Hopfield model with binary synapses when coding level $f=0.5$, to the model with fully connected or sparsely connected binary synaptic connectivity with arbitrary coding levels. Our results show that the network with binarized Hebbian rule has a similar capacity as the model with continuous synapses for any coding level,  and that this capacity tends asymptotically towards the optimal capacity obtained by Gutfreund and Stein \cite{Gutfreund}, in the space of all possible binary connectivity matrices. Our results also show that a network with sparse binary connectivity can have a larger information capacity per synapse than a fully connected network, and thus can allow a network to store information more efficiently.

\section{Results}
\subsection{Storage capacity of fully-connected network with binary synapses}
 We consider a fully-connected neural network with $N$ binary (0,1) neurons. The activity of neuron $i$ ($i = 1,...,N$) is described by a binary variable, $V_i = 0,1$. Each neuron is connected to other neurons through the connectivity matrix $W$. The activity of neuron $i$ at time $t$ is determined by the asynchronous update rule (see Appendix VI for details):

\begin{equation} \label{M:dy1}
\begin{aligned}
    V_i\left (t+1\right ) = \Theta[h_i\left (t\right ) - \theta],
\end{aligned}
\end{equation}
where
\begin{equation} \label{M:dy2}
\begin{aligned}
    h_i\left (t\right ) = \sum_{j \neq i}^N W_{i j}V_j\left (t\right ),
\end{aligned}
\end{equation}
is the local field, defined as the total input of neuron $i$, where $\theta$ is an activation threshold (constant independent of $V_i$), and $\Theta$ is the Heaviside function. \R{We also consider in Appendix a more general case of dynamics with stochastic updates characterized by a temperature $T$, but we focus in the main text in the zero temperature, deterministic limit.}

The storage capacity of the network whose dynamics is defined by Eq.(\ref{M:dy1}) and Eq.(\ref{M:dy2}) is determined by the connectivity matrix $W$. This connectivity matrix $W$ depends on $p$ random uncorrelated patterns $\Vec{\eta}^{\mu}$, $\mu = 1,...,p$, that are described by independent Bernoulli random variables:
\begin{equation} 
\begin{aligned}
    P\left (\eta_i^{\mu}\right ) = f\delta_{\left (\eta_i^{\mu},1\right )} + \left (1-f\right )\delta_{\left (\eta_i^{\mu},0\right )},
\end{aligned}
\end{equation}
where $\delta$ is the Kronecker delta function, and where $f$ is the coding level (the fraction of active neurons). The storage capacity $\alpha$ is defined as the maximal number of stored patterns $p$ divided by the network size $N$, $\alpha = p/N$.

In this paper, we construct connectivity matrix $W$ from the patterns $\Vec{\eta}^{\mu}$ using a `clipped' learning rule:
\begin{equation} \label{M:lr}
\begin{aligned}
        W_{i j} = \frac{\sqrt{p}}{N}F\left (\frac{1}{f(1-f)\sqrt{p}}\sum_{\mu=1}^{p}\left (\eta^{\mu}_i - f\right )\left (\eta^{\mu}_j - f\right )\right ),
\end{aligned}
\end{equation}
where $F$ is given by:
\begin{equation} \label{M:cf}
\begin{aligned}
        F\left (x\right ) = \sqrt{\frac{\pi}{2}} \mbox{sign}(x),
\end{aligned}
\end{equation}
where the prefactor $\sqrt{\pi/2}$ is used for convenience. Thus, $W_{i j}$ is only allowed to take two distinct values. With the nonlinear function $F(x)$ given by Eq.~(\ref{M:cf}), $W_{i j}$ can be positive or negative. In neurobiological networks, synaptic weights are sign-constrained, and their sign depends on whether the presynaptic neuron is excitatory or inhibitory. The network with the connectivity matrix given by Eqs.~(\ref{M:lr},\ref{M:cf}) leads to local fields of the form
\begin{equation}
    \label{M:hieq}
h_i = \sum_j W_{ij} V_j = \frac{\sqrt{2\pi p}}{N}\sum_j \Theta(x_{ij}) V_j  - \sqrt{\frac{\pi p}{2}} \frac{1}{N} \sum_j V_j,
\end{equation}
where $\Theta(x)$ is the Heaviside function, and $x_{ij}$ is the argument of $F$ in Eq.~(\ref{M:lr}). Eq.~(\ref{M:hieq}) shows that the network is equivalent to a purely excitatory network with binary weights (the first term in the r.h.s~of Eq.~(\ref{M:hieq})) with an instantaneous linear inhibition (the second term  in the r.h.s~of Eq.~(\ref{M:hieq})).

Notice also that when $F(x) = x$, Eq.(\ref{M:lr}) yields the learning rule in the model of Tsodyks and Feigel'man (TF) \cite{Tsodyks}, where $W_{i j}$ is a continuous variable. Therefore, we can interpret the learning rule of Eq.(\ref{M:lr}) as first learning patterns $\Vec{\eta^{\mu}}$ using the TF learning rule, and then clipping the weight into discrete values at the end of the learning phase. In the following, we call this model the Clipped Tsodyks-Feigel'man (CTF) model.

The nonlinearity of $F(x)$ in Eq.(\ref{M:cf}) makes the storage capacity more difficult to calculate than the one of a network with a linear learning rule. In 1986, Sompolinsky introduced a method to compute the storage capacity of Hopfield networks with non-linear learning rules \cite{Sompolinsky_1,Sompolinsky_2}. In particular he showed that in the large $N$ limit, these networks are equivalent to a linear learning rule with an added random Gaussian noise,
\begin{equation} \label{M:elr}
\begin{aligned}
    W_{ij} = \frac{J}{Nf(1-f)} \sum_{\mu}^{p}\left (\eta_i^{\mu} - f\right ) \left (\eta_j^{\mu} - f\right ) + \delta_{ij},
\end{aligned}
\end{equation}
where $J$ is an embedding strength, and $\delta_{i j}$ is a random symmetric Gaussian matrix. Both $J$ and the variance of the random Gaussian matrix $\Delta_0^2 = {N\langle \delta_{i j }^2\rangle}/({J^2}{\alpha})$ can be calculated as a function of $F(x)$ \cite{Sompolinsky_2}.  For $F(x)$ given by Eq.(\ref{M:cf}), the embedding strength $J$ and $\Delta_0^2$ are given by (see details in Appendix): 
\begin{equation} \label{M:JDelta0}
           J = 1,\quad \Delta^2_0 = \frac{\pi}{2} - 1.
\end{equation}

\subsubsection{\R{Calculation of the storage capacity for arbitrary coding level $f$}}

To compute the storage capacity of a network with a learning rule given by Eq.(\ref{M:elr}), we use standard methods and introduce the Hamiltonian
\begin{equation} \label{M:ham}
        H = \frac{1}{2}\sum_{i \neq j} W_{i j}V_i V_j + \theta \sum_{i}V_i,
\end{equation}
where $W_{i j}$ is given Eq.(\ref{M:elr}). The typical free energy of the system can be derived using the replica method \cite{Mezard,Amit,Sompolinsky_1}. The calculation allows us to derive order parameters characterizing the system (such as the overlap of network state with stored patterns), and its storage capacity. Using a replica symmetric ansatz, the free energy of the system can be characterized by five order parameters $m, Q, q, R, r$, where
\begin{equation} \label{M:op}
\begin{split}
        m & = \frac{1}{N} \sum_{i} \Tilde{\eta}_i^1 V_i,\\
        Q & = \frac{1}{N} \sum_i V_i,\\
        q & = \frac{1}{N} \sum_i V_i^2,
\end{split}
\end{equation}
and where $R$ and $r$ are conjugate variables of $Q$ and $q$,  and are defined by Eqs.(\ref{Rr}) in Appendix II. The order parameter $m$ measures the retrieval quality of a pattern stored in memory. Solutions with $\tilde{m} \equiv \frac{m}{f(1-f)} \sim 1$ represent `retrieval states' in which the network goes to a fixed point close to one of the stored patterns. Solutions with $\tilde{m} = 0$ corresponds to no retrieval. The order parameter $Q$ and $q$ represent the average neural activity and square of neural activity of the network (see Appendix II for more details).
The mean-field equations of the system are obtained using a saddle point method. The full equations are given by Eq.(\ref{full}) in Appendix II for arbitrary coding levels and temperature. \R{In the zero-temperature limit, the equations simplify to the following set of equations:}

\begin{equation} \label{0_T_main_text}
\begin{aligned}
        & \tilde{m} =\Phi(a_1) - \Phi(a_2),\\
        & \tilde{r} = f\Phi(a_1) + (1-f)\Phi(a_2),\\
        & a_1 = \frac{\tilde{\theta} - (1-f)\tilde{m} -Y}{\sqrt{\tilde{r} \alpha(1 + \Delta_0^2(1-C)^2)}},\\
        & a_2 = \frac{\tilde{\theta} + f\tilde{m} -Y}{\sqrt{\tilde{r} \alpha(1 + \Delta_0^2(1-C)^2)}},\\
        & Y = \frac{\alpha C f}{2 (1 - C)} + \frac{1}{2} \alpha C f \Delta_0^2,\\
        & C = \frac{f}{2\pi \alpha \tilde{r}} (f e^{-a_1^2/2} + (1-f)e^{-a_2^2/2})
\end{aligned}
\end{equation}
\R{where $\tilde{m} = m/f(1-f), \tilde\theta=\theta/f, \Delta_0^2 = \Delta^2/\alpha$ and $\Phi\left (x\right ) = \int_{x}^{\infty}Dz$.}

\begin{figure}[htbp]
\centering
\includegraphics[width=0.45\linewidth]{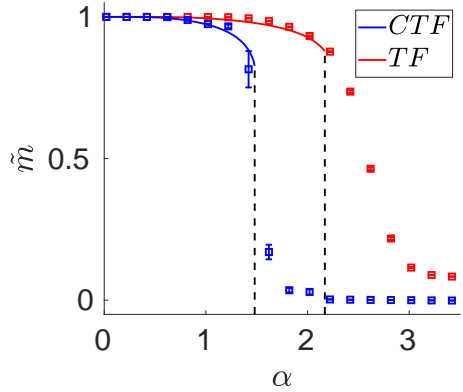}
\caption{Overlap as a function of storage load $\alpha$ for the Tsodyks-Feigel'man model (TF) and the clipped Tsodyks-Feigel'man model (CTF), for $f=0.02$. The solid lines represent the theoretical prediction and squares represent the simulation results with a network of size $N=4000$ (mean and standard deviation computed over five independent realizations). Dashed lines mark the storage capacity for CTF and TF. For both models, the neuronal activity threshold $\theta$ is chosen as the one that optimizes capacity. For the parameters chosen here ($f=0.02$), $ \tilde \theta \equiv \theta / f \approx 0.6$.}
\label{TM1} 
\end{figure}

\begin{figure}[htbp]
\centering
\includegraphics[width=0.75\linewidth]{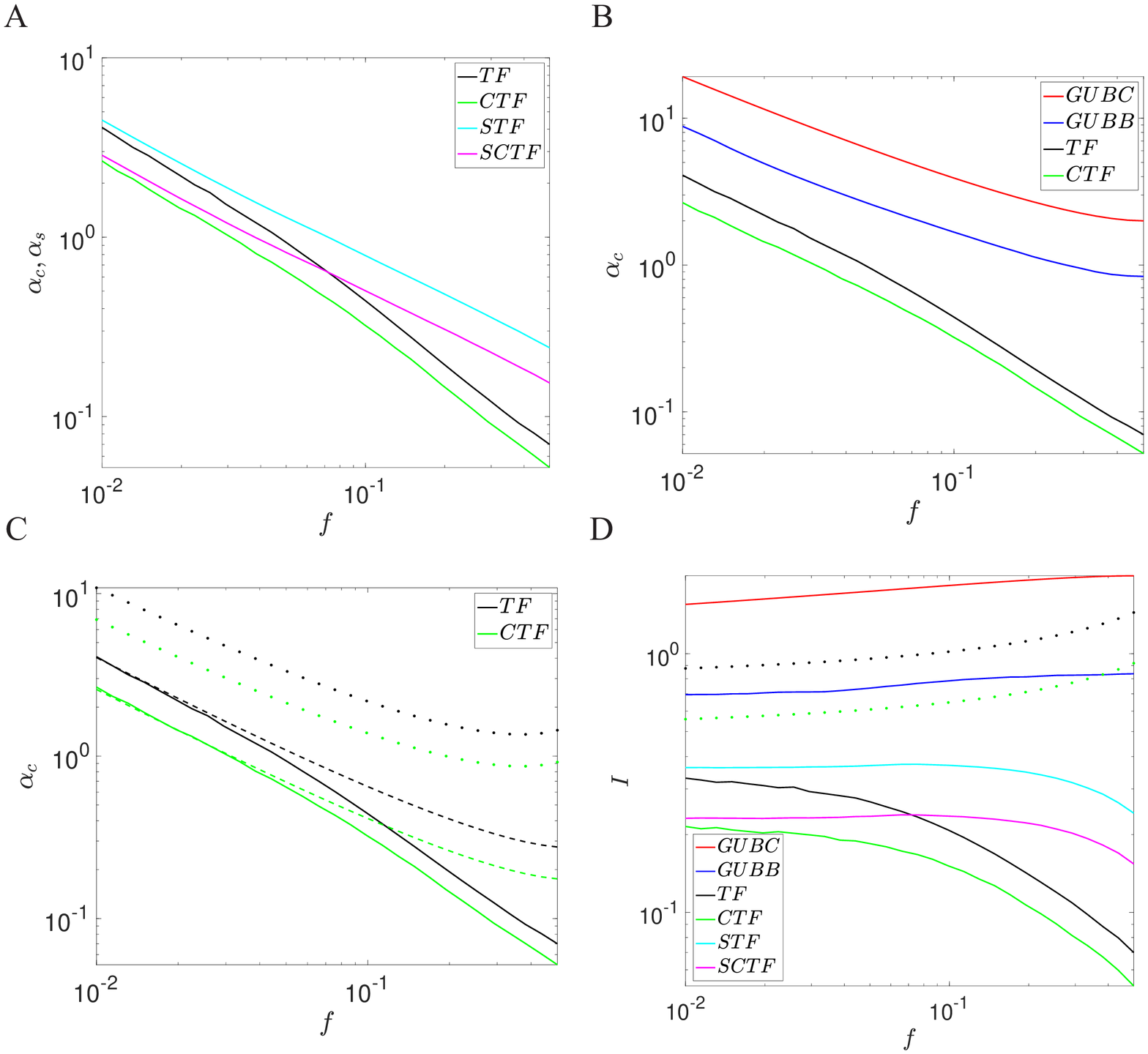}
\caption{Comparison between storage capacity of Hebbian rules and the respective upper bounds. GUBC: Gardner Upper Bound for networks with Continuous weights \cite{Gardner}. GUBB: Gutfreund and Stein Upper Bound for networks with Binary weights \cite{Gutfreund}. See more details about capacity upper bounds in Appendix IV. (A) Storage capacity of TF and CTF models as a function of coding level $f$. When $f \to 0$, both capacities increase as $1/f\ln(1/f)$. Cyan and magenta lines represent the storage capacity for a sparsely connected network with continuous weights (STF) and a network with binary weights (SCTF). The storage capacities of fully connected and sparsely connected networks converge when $f \to 0$. \R{Notice that for fully-connected network, storage capacity is defined as $\alpha_c = p/N$ while for the sparsely-connected network, storage capacity is defined as $\alpha_s = p/cN$, where $c \ll 1$.}(B) Comparison between storage capacity and upper bounds as a function of coding level $f$. Even at $f = 10^{-2}$, for both TF and CTF models, the capacity is only around a third of the respective bounds, and thus the asymptotic solution Eq.(\ref{M:mf}) is approached very slowly. (C) Comparison between the numerical solution and asymptotic solutions. Solid lines are the numerical solutions of TF and CTF models, the dotted lines with the same color are the corresponding asymptotic solutions in the sparse coding limit (Eq.(\ref{M:c})), and dashed lines represent asymptotic solutions with finite coding level corrections (Eq.(\ref{M:finite},\ref{M:finite_theta})). (D) Stored information per synapse as a function of coding level. When the coding level $f$ goes to 0, the information stored in synapses increases but with an extremely slow rate for both TF and CTF models. Dotted lines represents stored information of asymptotic solutions in the sparse coding limit (\R{i.e., $I(\alpha=(2f|\mbox{ln}f|)^{-1})$ for continuous weights case and $I(\alpha=(\pi f|\mbox{ln}f|)^{-1})$ for binary weights case).}}
\label{TM2} 
\end{figure}

Once $f, \alpha, \theta$ are given, Eq.(\ref{0_T_main_text}) can be solved numerically to obtain the order parameters, including the rescaled overlap with the retrieved pattern, $\tilde{m}$. Fig.\ref{TM1} shows $\tilde{m}$ as a function of $\alpha$ for both TF and CTF models for $f=0.02$. The figure shows that analytical results are in good agreement with simulations, using a network with $4,000$ neurons. The maximal capacity of the network $\alpha_c$ is given by the largest value of $\alpha$ for which there exist retrieval states (i.e. states with non-zero $\tilde{m}$), optimized over the threshold $\theta$. Fig.\ref{TM1} shows that the maximal capacity of the CTF model is lower than the one of the TF model, as expected, but only by a factor of about 1.5. This maximal capacity $\alpha_c$ is plotted as a function of $f$ in Fig.\ref{TM2}A. 

\subsubsection{\R{Sparse coding limit}}

In the biologically relevant sparse coding limit $f \to 0$, the mean-field Eq.(\ref{0_T}) take a rather simple form:
\begin{equation} \label{M:mf}
\begin{aligned}
        & \tilde{m} = \Phi\left (\frac{\tilde \theta - \tilde{m}(1-f)}{\sqrt{\tilde{r} \alpha  (1+\Delta^2_0 )}}\right ) - \Phi\left (\frac{\tilde \theta + \tilde{m} f}{\sqrt{\tilde{r} \alpha  (1+\Delta^2_0 )}}\right ),\\
        & \tilde{r} = f\Phi\left (\frac{\tilde{m}f}{\sqrt{\tilde{r} \alpha  (1+\Delta^2_0 )}}\right ) + (1-f) \Phi\left (\frac{\tilde \theta}{\sqrt{\tilde{r} \alpha  (1+\Delta^2_0 )}}\right ),\\
\end{aligned}
\end{equation}
where $\tilde \theta =\theta/f$ is a rescaled threshold, $\tilde{r}=r/f^2$, and $\Phi\left (x\right ) = \int_{x}^{\infty}Dz$ is the complementary cumulative distribution function of the standard Gaussian distribution.




With additional analysis (see details in Appendix III), we find that the maximum capacity in this limit is obtained when $\tilde \theta  \sim 1$, and the maximum capacity is:
\begin{equation} \label{M:c}
\begin{split}
        \alpha_c \simeq \frac{1}{\pi f \left|\log f \right|}.
\end{split}
\end{equation}
which can be compared with the capacity of TF model obtained by \cite{Tsodyks}, $\alpha_c\simeq 1/(2f|\log f|)$. Thus, the two capacities differ by a factor $\pi/2\sim 1.57$.
We next address the question of how close this capacity is to an upper bound in the space of all possible binary connectivity matrices. This upper bound was computed 
by Gutfreund and Stein \cite{Gutfreund} for arbitrary coding levels. The $f\rightarrow 0$ behavior could not be determined in a simple form in that paper, however it was shown that
the upper bound must be smaller or equal than $1/(\pi f|\log f|)$. Our asymptotic result, Eq.~(\ref{M:c}), indicates that the upper bound for binary connectivity matrices is indeed asymptotically $1/(\pi f|\log f|)$, and thus the clipped TF model becomes asymptotically optimal in the sparse coding limit. This is similar to what happens in networks with continuous synapses, for which it was shown that the storage capacity of the TF model tends to the upper bound of storage capacity obtained by Gardner \cite{Gardner}, in the space of all continuous synaptic matrices in the sparse coding limit. Thus, in spite of their remarkable simplicity, both  TF and CTF models provide close to optimal learning rules for models with continuous and discrete weights, respectively.  For convenience, we summarize the storage capacity of different models in Table 1 \cite{Krauth, Hopfield, Sompolinsky_1, Tsodyks, Gardner}.

\begin{table}
\begin{center}
\begin{tabular}{|l|c|c|}
     \hline
               &capacity of Hebbian rule & Upper Bound  \\
     \hline
     $f=0.5$, continuous $W_{i j}$ & $\sim 0.14$ & $2$ \\
     \hline
     $f=0.5$, binary $W_{i j}$ & $\sim 0.1$ & $\sim 0.83$ \\
     \hline
     $f \to 0$, continuous $W_{i j}$ & $(2f|\log f|)^{-1}$ & $(2f|\log f|)^{-1}$ \\
     \hline
     $f \to 0$, binary $W_{i j}$ & $(\pi f|\log f|)^{-1}$ & $(\pi f|\log f|)^{-1}$ \\
     \hline
\end{tabular}
\end{center}
\caption{Comparison between the storage capacity of Hebbian rules and upper bounds computed using Gardner approach.}

\end{table}

\subsubsection{\R{Leading correction to sparse coding limit}}
We can see that the capacities of networks with Hebbian rules in the unbiased case ($f = 0.5$) are much smaller than the corresponding upper bounds, while they converge to the corresponding upper bounds in the sparse coding limit. However, solving numerically mean-field Eqs.(\ref{0_T}) for finite coding level $f$, we find that the capacities of TF and CTF converge to the upper bounds extremely slowly (see Fig.~\ref{TM2}B). As shown in Fig.\ref{TM2}B, the capacity of Hebbian rules is only around $1/3$ compared to their corresponding upper bounds when the coding level $f = 10^{-2}$. This is mainly because the optimal threshold $\tilde \theta$ approaches 1 extremely slowly when $f$ decreases (for instance, for $f=0.02$, $\tilde \theta\sim 0.6$). With additional analysis (see Appendix III), we derived the leading correction to the asymptotic solution at the finite coding level :
\begin{equation} \label{M:finite}
\begin{split}
        \alpha_c \simeq \frac{\tilde{\theta}^2_{opt}}{\pi f |\log f|},
\end{split}
\end{equation}
where the optimal threshold $\tilde{\theta}_{opt}$ is obtained by solving the equation
\begin{equation} \label{M:finite_theta}
\begin{split}
        \frac{2\tilde{\theta}_{opt}^2 \left|\log(1-\tilde{\theta}_{opt})\right|}{(1-\tilde{\theta}_{opt})^2} = \left|\log f \right|.
\end{split}
\end{equation}
Notice that when coding level $f \to 0$, $\tilde{\theta}_{opt} \to 1$ and Eq.(\ref{M:finite}) recovers the asymptotic scaling of Eq.(\ref{M:c}).
The asymptotic solutions Eq.(\ref{M:c}) and Eq.(\ref{M:finite}) are compared in Fig.\ref{TM2}C, and we can see that Eq.(\ref{M:finite}) agrees with the numerical solutions very well when the coding level $f$ is small. This result indicates that, in the biological sparse coding limit (i.e., coding level is small but finite), the capacities of models with Hebbian rules are still notably smaller than the maximal capacities, in the space of all possible connectivity matrices. 

While the storage capacity in terms of numbers of patterns stored per synapse diverges in the sparse coding limit, the information stored per pattern decreases in that limit since it is proportional to the binary entropy of $f$. As a result, the total information stored per synapse remains finite in the sparse coding limit, both in the TF model \cite{Tsodyks} and in the corresponding Gardner bound \cite{Gardner}.
Fig.\ref{TM2}D shows the information capacity in bits per synapse for different models as a function of $f$,
\begin{equation} \label{M:I}
\begin{split}
        I = - \frac{\alpha}{\ln 2} \left ( f \ln f + (1-f) \ln (1-f) \right)
\end{split}
\end{equation}
We find that when the coding level $f$ decreases, the information capacity of TF and CTF increases quickly, while the corresponding upper bounds decrease slowly. When $f$ goes to 0, the information capacity $I$ of TM and CTM further increase and eventually converge to the optimal information capacity, but the convergence rate is extremely low.

\subsection{Storage capacity of sparsely-connected network with binary synapses}

Cortical networks are characterized by low connection probabilities between neurons (e.g.~\cite{Markram}). In the case we interpret the low synaptic efficacy state to be zero, the network we have studied so far has a 50\% connection probability, much higher than observed connection probabilities in cortex, which are of order 10\% for excitatory neurons at short distances ($<100 \mu$m). This motivates the study of networks with sparser connectivity. Here, we study two cases - one in which sparse connectivity is uncorrelated with learning, and the other one where sparse connectivity is an outcome of learning with a high synaptic threshold.

\subsubsection{Sparse connectivity uncorrelated with learning}

We first consider the case where learning occurs on top of a sparse random Erdos-Renyi `structural' connectivity matrix,
\begin{equation} \label{M:slr}
\begin{aligned}
        W_{i j} = \frac{c_{ij}\sqrt{p}}{Nc}F\left (\frac{1}{f(1-f)\sqrt{p}}\sum_{\mu=1}^{p}\left (\eta^{\mu}_i - f\right )\left (\eta^{\mu}_j - f\right )\right ),
\end{aligned}
\end{equation}
where $F(x)$ is the same as the clipped function Eq.(\ref{M:cf}) for fully-connected case, 
and $c_{ij}=1,0$ is a random binary matrix, with 
\begin{equation} 
\begin{aligned}
    P\left (c_{ij}\right ) = c\delta_{\left (c_{ij},1\right )} + \left (1-c\right )\delta_{\left (c_{ij},0\right )},
\end{aligned}
\end{equation}
where $0<c\ll1$ is the connection probability. 
The storage capacity of learning rule Eq.(\ref{M:slr}) can be calculated similarly as the model in \cite{Tsodyks_2} in the sparse connectivity limit $c\ll 1$, and the mean-field equations for finite coding level $f$ are given as (see details in Appendix V): 
\begin{equation} \label{M:smf}
\begin{split}
       &\tilde{m} = \Phi \left(  \frac{\tilde \theta - \tilde{m}(1-f)}{\sqrt{\alpha_s q(1+ \Delta_0^2)}}\right) - \Phi \left(  \frac{\tilde \theta + \tilde{m}f}{\sqrt{\alpha_s q(1+ \Delta_0^2)}}\right),\\
       &q = f\Phi \left(  \frac{\tilde \theta - \tilde{m} (1-f)}{\sqrt{\alpha_s q (1+ \Delta_0^2)}}\right) + (1-f)\Phi \left(  \frac{\tilde \theta + \tilde{m}f}{\sqrt{\alpha_s q(1+ \Delta_0^2)}}\right),
\end{split}
\end{equation}
where $\alpha_s = p/cN$ and where other order parameters are defined in Eq.(\ref{M:op}). The numerical solution of Eq.(\ref{M:smf}) are compared with fully-connected case in Fig.\ref{TM2}A. In the sparse coding case, the mean-field Eq.(\ref{M:smf}) coincide with the mean-field Eq.(\ref{M:mf}), as expected \cite{Tsodyks_2}. We see that the capacity, in terms of number of patterns stored divided by number of connections per neuron, is larger in the sparsely connected case than in the fully connected case, as expected from previous results in networks with continuous synapses \cite{Derrida,Tsodyks_2}.


\subsubsection{Sparse connectivity induced by learning}

In this section, we consider the case where sparse connectivity is obtained by adding a threshold to the clipped function. Here we generalize the clipped function Eq.~(\ref{M:cf}) to:
\begin{equation} \label{M:gcf}
\begin{split}
      F_T(x) = \sqrt{2 \pi}(\Theta(x-T) - M),
\end{split}
\end{equation}
where M is given by
\begin{equation} 
\begin{split}
      M  = \frac{1}{\sqrt{2 \pi}}\int_{-\infty}^{\infty}\Theta_T(x)e^{-x^2/2}d x,
\end{split}
\end{equation}
where $\Theta_T(x) = \Theta(x-T)$, and $T$ is the threshold that can be used to increase the sparseness of network connectivity, interpreting the low synaptic state as a 0 state. 
With such a connectivity matrix, the connection probability $R_1$ is given by:
\begin{equation} \label{M:R1}
\begin{split}
      R_1 = \frac{1}{2}\left(1-\mbox{erf}(\frac{T}{\sqrt{2}})\right)
\end{split}
\end{equation}

\begin{figure}[htbp]
\centering
\includegraphics[width=0.75\linewidth]{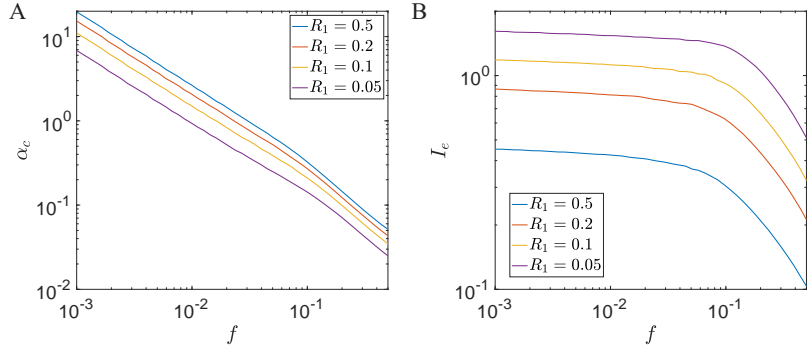}
\caption{Storage capacity and information capacity for a network with sparse excitatory binary connections. (A) Storage capacity as a function of connection probability $R_1$ and coding level $f$. The storage capacity decreases when $R_1$ decreases. (B) Information capacity per active synapse, as a function of $f$ and $R_1$. The information per synapse $I_e$ increases when $f$ and $R_1$ decrease. This indicates that for learning rule Eqs.(\ref{M:lr},\ref{M:gcf}), both sparse coding and sparse connectivity can improve the coding efficiency of the network. This result also indicates that the network can have an optimal $R_1$ to balance storage capacity and coding efficiency. }
\label{TM3} 
\end{figure}

In this case, the embedding strength $J$ and additional noise introduced by clipped function Eq.(\ref{M:gcf}) $\Delta_0^2 = {N\langle \delta_{i j }^2\rangle}/{J^2}{\alpha}$  are
\begin{equation} 
\begin{split}
       J = e^{-T^2}, \Delta_0^2 = \frac{\pi}{2}e^{T^2}\left(1-\mbox{erf}\left(\frac{T}{\sqrt{2}}\right)^2\right)-1
\end{split}
\end{equation}
Notice that the noise strength $\Delta_0^2 = {N\langle \delta_{i j }^2\rangle}/{J^2}{\alpha}$ is an increasing function of $T$. The storage capacity of the learning rule Eq.(\ref{M:lr},\ref{M:gcf}) is determined by mean-field Eq.(\ref{M:mf}) for a given connection probability $R_1$ and coding level $f$. As shown in Fig.\ref{TM3}A, the storage capacity $\alpha_c$ decreases when the threshold increases, and consequently the connection probability $R_1$ decreases. Unsurprisingly, the storage capacity decreases as the fraction of non-zero synapses decreases. However, storing information with a smaller number of synapses also carries benefits in terms of efficiency of information storage.

\begin{figure}[htbp]
\centering
\includegraphics[width=0.8\linewidth]{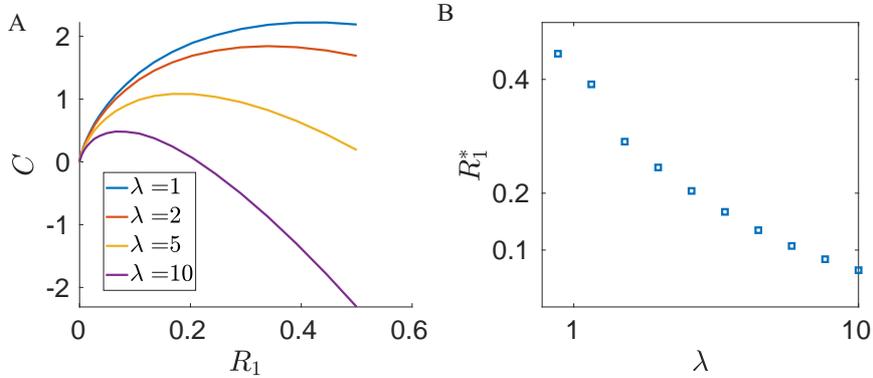}
\caption{\R{Connection probability that optimizes capacity subject to a synapse maintenance cost $\lambda$. (A) Cost function $C$ as a function of $R_1$ and $\lambda$. (B) Optimal connection probability $R_1^*$ for different $\lambda$. We can see that the more costly synaptic maintenance is, the sparser the resulting connectivity. In both (A) and (B), coding level is set to be $f=0.01$.} }
\label{TM4} 
\end{figure}

To quantify this efficency, we calculate the information stored in the network per non-zero synapse, $I_e$:
\begin{equation} \label{M:eI}
\begin{split}
        I_e = - \frac{\alpha}{R_1\ln 2} \left ( f \ln f + (1-f) \ln (1-f) \right)
\end{split}
\end{equation}
The relation between $I_e$, $R_1$ and $f$ is shown in Fig.\ref{TM3}B. We see that the information capacity per synapse increases when the excitatory connectivity becomes more sparse. This is mainly because we only keep connections for which the Hebbian term (i.e.~the argument of the clipping function $F$ in Eq.~(\ref{M:gcf}) is large. In this way, the network can encode information more efficiently when excitatory connections become sparse. \R{Note that maximizing storage capacity subject to a constraint of minimizing the fraction of active synapses would lead to an optimal connection probability $R_1^*$, whose precise value would depend on the cost of maintenance of an active synapse. One can define a cost function $C = \alpha - \lambda R_1$, where second term represents the cost of maintenance of active synapses. For a given $\lambda$, we can obtain $R_1^*$ that maximize $C$. As shown in Fig.\ref{TM4}B, the more costly synaptic maintenance is, the sparser the resulting connectivity. And the optimal connection probability $R_1^* \sim 0.1$ when $\lambda \sim 10$. }
\FloatBarrier

\FloatBarrier

\section{Discussion}

We have calculated the storage capacity of an attractor neural network endowed with binary synaptic weights at arbitrary coding levels. Our results show that a network with a binarized Hebbian learning rule has a capacity that is close to the capacity of a network with continuous weights at any coding level, since the decrease in capacity is only about 1.5 compared to continuous weights. Our results generalize the results obtained by Sompolinsky for a coding level $f = 0.5$ \cite{Sompolinsky_2}, to arbitrary coding levels. Furthermore,  our analysis shows that the storage capacity of CTF tends in the sparse coding limit to the upper bound of storage capacity, in the space of all possible binary connectivity matrices.  We also provide a finite coding level correction for this asymptotic solution, and the results indicate the capacities of TF and CTF converge extremely slowly to the optimal capacity when the coding level decreases, since the corrections are of order $1/\sqrt{\log(1/f)}$. In particular, for $f=0.01$ \cite{Waydo}, the capacity of the clipped model is only about a third of the upper bound. Our results also show that sparse connectivity matrices can allow these networks to have a larger information capacity per synapse and thus encode information more efficiently. 

The binary connectivity matrices used in this paper were constructed using a clipped function whose argument is an analog variable containing information about all stored patterns. This assumes that the synapse can store continuous information during the learning phase, before binarizing this information. An alternative scenario is that the synapse is required to be discrete during all learning phases. Tsodyks, Amit and Fusi studied models with discrete synapses under an online learning setting in which synapses only have information about the currently shown pattern to make a transition between states. They showed that this leads to a drastic decrease in storage capacity when the coding level is $f  = 0.5$ \cite{Tsodyks_3, Amit_Fusi_1, Amit_Fusi_2}, since in that case the total number of stored patterns can scale at most as $\sqrt{N}$, implying a vanishing amount of information stored per synapse in the large $N$ limit. Later work found that a storage capacity of order 1 bit/synapse can be recovered in the sparse coding limit ($f\sim \log(N)/N$), even when synapses are required to be discrete during all phases of learning \cite{Amit_Fusi_2, Dubreuil}.

Another scenario studied by multiple authors consists in synapses with binary weights with multiple hidden states \R{(describing e.g.~different configurations of protein interaction networks on the post-synaptic side)} \cite{Fusi2005,Fusi2007,Ganguli,Benna_Fusi}. With appropriate structure of hidden states, such synapses can greatly extend the time for which synaptic connectivity can remain correlated with a pattern shown at a particular time. This scenario has been primarily studied using a signal-to-noise analysis quantifying the degree of correlation of the synaptic matrix with patterns presented to the network. To our knowledge, this scenario has never been implemented in attractor network models, and thus the storage capacity in these multi-state models is still an open question. More experimental data will be necessary to understand which class of models best captures synaptic plasticity in neurobiological synapses.






\section*{Appendix}

\subsection*{I. Calculation of $J$ and $\Delta_0$}

To compute $J$ and $\Delta_0$ in Eq.~(\ref{M:JDelta0}) we use the same strategy as Sompolinsky \cite{Sompolinsky_1,Sompolinsky_2}.
We first calculate the average overlap between a given pattern and the local field when the network is retrieving that pattern. Let us denote $\Tilde{\eta}_i^{\mu} = \eta_i^{\mu} - f$. For the learning rule given by Eq.~(\ref{M:elr}), we have:
\begin{equation} \label{o1}
\begin{split}
     \langle \Tilde{\eta}^1_i\sum_j \Tilde{\eta}^1_j W_{ij}\rangle
      = J f \left (1-f\right ) 
\end{split}
\end{equation}
Similarly, the average overlap for clipped learning rule Eq.~(\ref{M:lr}) is:
\begin{equation} \label{o2}
\begin{split}
     \left \langle \Tilde{\eta}^1_i\sum_j \Tilde{\eta}^1_j W_{i j} \right \rangle
      &= \frac{\sqrt{p}}{N} \left \langle \sum_j \Tilde{\eta}^1_i\Tilde{\eta}^1_j F\left (\frac{\sum_{\mu}^{p}\Tilde{\eta}^{\mu}_i\Tilde{\eta}^{\mu}_j}{\sqrt p{f\left (1-f\right )}}\right ) \right \rangle\\
      &= f \left (1-f\right ) \left \langle  x F\left (x\right ) \right \rangle\\.
\end{split}
\end{equation} where 
\begin{equation} \label{cl}
\begin{split}
    x = \frac{\sum_{\mu}^{p}\Tilde{\eta}^{\mu}_i\Tilde{\eta}^{\mu}_j}{\sqrt{p}{f\left (1-f\right )}} .
\end{split}
\end{equation}
In the large $p$ limit,  $x$ becomes a random variable drawn from a standard Gaussian distribution, $x \sim N\left (0,1\right )$ according to the Central Limit Theorem. Thus, from Eqs.~(\ref{o1},\ref{o2}), we obtain the embedding strength $J$ as:
\begin{equation}\label{strength}
    J = \langle  xF\left (x\right )\rangle,
\end{equation}

In order to obtain the variance $\delta_{i j}$ of Eq.~(\ref{M:elr}), we calculate the variance of synaptic weights for both linear and clipped learning rules. In the linear case, we have:
\begin{equation} \label{t1}
\begin{split}
        N^2\langle W_{i j}^2\rangle &=  N^2 \frac{J^2}{ f^2\left (1-f\right )^2 N^2} \left \langle \left (\sum_\mu^p \Tilde{\eta}_i^\mu \Tilde{\eta}_j^\mu\right )^2 \right \rangle\\
        &=  p J^2 \left \langle \left (\sum_\mu^p \frac{\Tilde{\eta}_i^\mu \Tilde{\eta}_j^\mu}{f\left (1-f\right )\sqrt p}\right )^2 \right\rangle\\
        &=  N \alpha J^2.
\end{split}
\end{equation}
The $<\cdot>$ in Eq.~(\ref{t1}) goes to 1 when $p \to \infty$. For the clipped $F\left (x\right )$, the variance of the weights is:
\begin{equation}\label{t2}
\begin{split}
        N^2\langle W_{i j}^2\rangle &= N^2 \frac{1}{N^2}p \left \langle F^2\left (\sum_\mu^p \frac{\Tilde{\eta}_i^\mu \Tilde{\eta}_j^\mu}{f\left (1-f\right )\sqrt p}\right ) \right \rangle\\
        &= p \langle F^2\left (x\right )\rangle\\
        &= N\alpha \tilde{J}^2,
\end{split}
\end{equation}
where we denote $\langle F^2\left (x\right )\rangle$ as $\Tilde{J}^2$. From Eqs.~(\ref{t1},\ref{t2}), we can see the additional noise introduced by a nonlinear $F\left (x\right )$ is:
\begin{equation}
        \langle  \delta_{i j }^2 \rangle = \frac{\alpha}{N} \left (\Tilde{J}^2-J^2\right ).
\end{equation}
Let $\Delta^2_0$ denote ${N\langle \delta_{i j }^2\rangle}/ {J^2}{\alpha}$, we have:
\begin{equation} \label{noise}
           \Delta^2_0 =  \left (\frac{\tilde{J^2}}{J^2}-1\right ).
\end{equation}
For $F\left (x\right )$ given by Eq.~(\ref{M:cf}), we obtain the embedding strength and noise parameter as 
\begin{equation} \label{para}
           J = 1, \quad \Delta^2_0 = \pi/2 - 1.
\end{equation}

\subsection*{II. Storage capacity of the fully-connected CTF model}

The Hamiltonian for the learning rule (\ref{M:elr}) is
\begin{equation} \label{ham}
        H = \frac{1}{2}\sum_{i \neq j} W_{i j}V_i V_j + \theta \sum_{i}V_i,
\end{equation}
where $W_{i j}$ is determined from  Eqs.(\ref{M:elr},\ref{para}). We can calculate the corresponding free energy by using the replica method (see e.g.~\cite{Amit,Tsodyks}). To compute the average logarithm of the partition function over the distribution of all random binary patterns $\langle \log Z \rangle$ directly, we can use the relation:
\begin{equation}
       \langle  \log Z \rangle = \lim\limits_{n \to 0} \frac{\langle Z^n\rangle-1}{n}.
\end{equation}
For the Hamiltonian in equation (\ref{ham}), we have:
\begin{equation}
\begin{split}
        \langle  \langle  Z^n \rangle \rangle \propto & \Big \langle  \Big \langle  \mbox{Tr}_{V^a} \exp\left (\frac{\beta J}{f (1-f) 2N} \sum_{\mu a} (\sum_{i} \Tilde{\eta}_i^\mu V_i^a )^2 
                        + \frac{\beta}{2} \sum_{i j a} \delta_{i j} V_i^a V_j^a - \beta \theta \sum_{i a} V_i^a\right )\Big \rangle \Big \rangle  ,
\end{split}
\end{equation}
where $a=1,\ldots,n$ is the replica index and the double brackets mean the average over both $V_i^a$ and $\eta_i^{\mu}$. We are  interested in the overlaps between the network state and the patterns stored in memory. We assume that the network has a macroscopic overlap with a single stored pattern ($\mu=1$ the one currently being retrieved by the network) and define the following order parameters: 
\begin{equation}
\begin{split}
        m_a & \equiv \frac{1}{N} \sum_{i} \Tilde{\eta}_i^1 V_i^a,\\
        m_{\mu a} &\equiv \frac{1}{\sqrt{N}} \sum_{i} \Tilde{\eta}_i^\mu V_i^a.
\end{split}
\end{equation}
The partition function can be rewritten in terms of these order parameters as
\begin{equation} \label{Z}
\begin{split}
        \left \langle \left \langle Z^n \right \rangle \right \rangle  \propto  & \Big \langle \Big \langle Tr_{V^a} \int  \left (\prod_{\mu a} d m_a^{\mu}\frac{\beta N}{2 \pi}\right )\\
               \times & \exp  \Big(  -\frac{\beta J N}{2 f(1-f)} \sum_{a} m_a^2 + \frac{\beta J}{f (1-f) } \sum_a m_a \sum_i \Tilde{\eta}_i^1 V_i^a\\
           - & \frac{\beta J}{2 f (1-f)} \sum_{\mu a} \left (m_a^{\mu}\right)^2 + \frac{\beta J}{\sqrt{N} f(1-f)}\sum_{\mu a} m_a^{\mu}\sum_{i} \tilde{\eta}_i^{\mu} V_i^a\\
           + & \frac{\beta}{2} \sum_{i j a} V_i^a V_j^a \delta_{i j} - \beta \theta \sum_{i a} V_i^a \Big) \Big \rangle \Big \rangle.
\end{split}
\end{equation}
The terms including $\eta_i^\mu$ and $\delta_{i j}$ in equation (\ref{Z}) can be averaged,
\begin{equation} \label{1}
\begin{split}
         \Big \langle \exp\left(\frac{\beta J}{\sqrt{N} f(1-f)}\sum_{\mu a} m_a^{\mu}\sum_{i} \tilde{\eta}_i^{\mu} V_i^a\right) \Big \rangle\ \propto\    \exp\left (\frac{\beta^2 J^2}{2N f (1-f)}\sum_{i \mu a b} V_i^a V_j^a m_a^{\mu} m_b^{\mu}\right ),
\end{split}
\end{equation}
\begin{equation} \label{2} 
\begin{split} 
         \Big \langle \exp\left ( \frac{\beta}{2} \sum_{i j a} V_i^a V_j^a \delta_{i j}\right ) \Big \rangle\ \propto\ 
         \exp\left (N \beta^2 J^2 \Delta^2 \sum_{a b} \left (\frac{1}{N} \sum_i V_i^a V_i^b \right )\left (\frac{1}{N} \sum_i V_i^a V_i^b \right ) \right ),
\end{split}
\end{equation}
where $\Delta^2 = {N\langle \delta^2_{i j}\rangle}/{ J^2}$. We then introduce the order parameters:
\begin{equation} \label{3}
\begin{split}
        Q_a &= \frac{1}{N} \sum_i V_i^a,\\
        q_{a b} &= \frac{1}{N} \sum_i V_i^a V_i^b,
\end{split}
\end{equation}
and use the integral representation of $\delta$ function :
\begin{equation} \label{4}
\begin{split}
        \delta\left (N Q_a - \sum_i V_i^a\right ) &= \int \frac{d R_a}{2 \pi} e^{-R_a\left (N Q_a - \sum_i V_i^a\right )},\\
        \delta\left (N  q_{a b} - \sum_i V_i^a V_i^b\right ) &= \int \frac{d r_{a b}}{2 \pi}e^{-r_{ab}\left (N q_{a b} - \sum_i V_i^a V_i^b\right )}.
\end{split}
\end{equation}
Combining Eqs.~(\ref{Z},\ref{1},\ref{2},\ref{3},\ref{4}), the partition function can be written as:
\begin{equation} \label{Rr} 
\begin{split}
        \langle \langle Z^n\rangle\rangle \propto \int \left (\prod_{a}dQ_a dR_a\right ) \left (\prod_{a \langle  b} q_{a b} r_{a b}\right ) e^{- N \beta g\left (\beta,m,Q,q,R,r\right )},
\end{split}
\end{equation}
where
\begin{equation} \label{f}
\begin{split}
        g = &\frac{J}{2f(1-f)} \sum_{a}m_a^2 + \theta \sum_a Q_a - \frac{1}{\beta} \sum_a R_a Q_a - \beta J^2 \Delta^2 \sum_{a b} q_{a b}^2 - \frac{1}{\beta} \sum_{a\langle b} r_{ab}q_{ab}\\
        & - \frac{1}{\beta} \log Tr_V^a \exp\left (\frac{\beta J}{f (1-f)} \sum_a \tilde{\eta}^1 m_a V^a - \sum_a R_a V^a - \sum_{a \langle  b}r_{ab}V^a V^b\right ) \\
        & - \frac{\alpha}{\beta} \log  \int \left (\prod_a d m_a\right ) \exp\left (-\frac{\beta J}{2 f (1-f)}\sum_a m_a^2 + \frac{\beta^2J^2}{2 f(1-f)}\sum_{a\neq b}m_a m_b q_{ab}\right ).
\end{split}
\end{equation}
The free energy per neuron is given as:
\begin{equation} \label{pf}
\begin{split}
        G/N = \lim\limits_{n \to 0} \frac{1}{n} \mbox{min}\  g\left (\beta, m, Q, q, R, r\right ).
\end{split}
\end{equation}
In the large $N$ limit, min $g\left (\beta, m, Q, q, R, r\right )$ is dominated by its value at saddle points. Next we will give the saddle point equations using replica symmetric $ansatz$. We assume that saddle-point values of the order parameters are not dependent on their replica index:
\begin{equation}
\begin{split}
        &m_a = m,\\
        &Q_a = Q, R_a = R,\\
        &q_{a b} = q, r_{a b} = r\ \left (a \neq b\right ).
\end{split}
\end{equation}
Now the free energy per neuron is simplified to:
\begin{equation} \label{res}
\begin{split}
        G/N &= \frac{m^2}{2f(1-f)} + \frac{\alpha}{2\beta} \{log\left (1 - \beta J \left (Q-q\right )\right ) - \frac{\beta J q}{1-\beta J \left (Q-q\right )} \} \\
        &-\frac{r q}{2\beta} + \frac{R Q}{\beta} + \theta Q - \frac{1}{4}\beta J^2 \Delta^2\left (q^2-Q^2\right )\\
        &-\frac{1}{\beta}\int Dz \log\left (1+\exp\left (\frac{\beta J}{f(1-f)} m \tilde{\eta} + R - \frac{r}{2} + \sqrt{r}z\right )\right ).
\end{split}
\end{equation}
The saddle point equations are obtained by setting the derivatives of $G/N$ to 0:
\begin{equation} \label{full}
\begin{aligned}
    &\frac{m}{f(1-f)} = \int Dz \Big \langle \Big \langle \tilde{\eta}K\left (\beta\left ( \frac{J m \tilde{\eta} }{f(1-f)}+ \frac{1}{\beta} \left (R-\frac{\alpha \beta^2 \tilde{r} + \beta^2 J^2 \Delta^2  q}{2}\right ) +\sqrt{\alpha \tilde{r} + J^2 \Delta^2 q }z\right )\right ) \Big \rangle \Big \rangle,\\
    &R-\frac{\alpha \beta^2 \tilde{r} + \beta^2 J^2 \Delta^2 q}{2} = \frac{\alpha}{2} \frac{\beta (Q-q)J}{1-\beta J \left (Q-q\right )} - \beta \theta + \frac{1}{2} \beta^2 \left (Q-q\right ) J^2 \Delta^2,\\
    &\tilde{r} = \frac{ J^2 q}{\left (1- J \beta \left (Q-q\right )\right )^2},\\
    &Q = \int Dz \Big \langle \Big \langle K\left (\beta\left ( \frac{J m \tilde{\eta}}{f(1-f)} + \frac{1}{\beta} \left (R-\frac{\alpha \beta^2 \tilde{r} + \beta^2 q}{2}\right ) +\sqrt{\alpha \tilde{r} + J^2 \Delta^2 q}z\right )\right ) \Big \rangle \Big \rangle,\\
    &q = \int Dz \Big \langle \Big \langle K^2\left (\beta\left ( \frac{J m \tilde{\eta}}{f(1-f)} + \frac{1}{\beta} \left (R-\frac{\alpha \beta^2 \tilde{r} + \beta^2 q}{2}\right ) +\sqrt{\alpha \tilde{r} + J^2 \Delta^2 q}z\right )\right ) \Big \rangle \Big \rangle,
\end{aligned}
\end{equation}
where $\tilde{r} = \frac{1}{\beta^2 \alpha}\left (r - \beta^2 J^2 \Delta^2 q\right )$, $K\left (x\right ) = \left (1+\exp\left (-x\right )\right )^{-1}$, and $Dz = dz \frac{\exp\left (-x^2/2\right )}{\sqrt{2\pi}}$. 
In the zero temperature limit $\beta \to \infty$, these saddle points equations can be simplified to:
\begin{equation} \label{0_T}
\begin{aligned}
        & \tilde{m} =\Phi(a_1) - \Phi(a_2),\\
        & \tilde{r} = f\Phi(a_1) + (1-f)\Phi(a_2),\\
        & a_1 = \frac{\tilde{\theta} - (1-f)\tilde{m} -Y}{\sqrt{\tilde{r} \alpha(1 + \Delta_0^2(1-C)^2)}},\\
        & a_2 = \frac{\tilde{\theta} + f\tilde{m} -Y}{\sqrt{\tilde{r} \alpha(1 + \Delta_0^2(1-C)^2)}},\\
        & Y = \frac{\alpha C f}{2 (1 - C)} + \frac{1}{2} \alpha C f \Delta_0^2,\\
        & C = \frac{f}{2\pi \alpha \tilde{r}} (f e^{-a_1^2/2} + (1-f)e^{-a_2^2/2})
\end{aligned}
\end{equation}
where $\tilde{m} = m/f(1-f), \tilde\theta=\theta/f, \Delta_0^2 = \Delta^2/\alpha$ and $\Phi\left (x\right ) = \int_{x}^{\infty}Dz$.

\subsection*{III. Sparse coding limit}

In the sparse coding limit $f \to 0$, $C$ goes to zero and $Y$ goes to zero. Eqs.~(\ref{0_T}) become:
\begin{equation} 
\begin{aligned}
        & \tilde{m} = \Phi\left (\frac{\tilde \theta - \tilde{m}(1-f)}{\sqrt{\tilde{r} \alpha \left (1+\Delta^2_0\right )}}\right ) - \Phi\left (\frac{ \tilde \theta+f\tilde{m}}{\sqrt{\tilde{r} \alpha \left (1+\Delta^2_0\right )}}\right ),\\
        & \tilde{r} = f\Phi\left (\frac{ \tilde \theta - f\tilde{m}}{\sqrt{\tilde{r} \alpha \left (1+\Delta^2_0\right )}}\right ) + \Phi\left (\frac{\tilde \theta}{\sqrt{\tilde{r} \alpha \left (1+\Delta^2_0\right )}}\right ),\\
\end{aligned}
\end{equation}
In the small $f$ limit, $\tilde{m} \sim 1$ requires:
\begin{equation} \label{approx1}
\begin{split}
       &\Phi\left (\frac{\tilde \theta - \tilde{m}(1-f)}{\sqrt{\alpha f \pi /2}}\right ) \sim 1, \quad \Phi\left (\frac{\tilde \theta+\tilde{m}f}{\sqrt{\alpha f \pi /2}}\right ) \ll 1
\end{split}
\end{equation}
Furthermore, $\alpha$ is maximized when $\tilde{r}$ is minimized, which requires the stronger condition
\begin{equation} \label{approx2}
\begin{split}
       &\Phi\left (\frac{\tilde \theta+\tilde{m}f}{\sqrt{\alpha f \pi /2}}\right ) \ll f,
\end{split}
\end{equation}
which leads to $\tilde{r}\sim f$.
Using $\lim\limits_{x \to +\infty} \Phi\left (x\right ) \simeq \frac{1}{x\sqrt{2\pi}}\exp\left (\frac{-x^2}{2}\right )$, in the small $f$ limit, Eq.(\ref{approx2}) gives:
\begin{equation}\label{ueq1}
\begin{split}
       \sqrt{\frac{\alpha f}{2 \tilde \theta^2}}\exp\left(-\frac{\tilde \theta^2}{\pi f\alpha}\right) \ll f.
\end{split}
\end{equation}

Rewriting $\alpha={k} / {f \log\left (f^{-1}\right )}$, we find that Eq.~(\ref{ueq1}) is satisfied provided $k < {\theta^2}/{\pi}$. Thus, the maximum storage capacity $\alpha$ increase with $\tilde \theta^2$ as:
\begin{equation} \label{s3}
\begin{split}
        \alpha_c \simeq \frac{\tilde \theta^2}{\pi f |\log f|}.
\end{split}
\end{equation}
In the sparse coding limit, the optimal threshold is obtained at $\tilde \theta = 1$ (the maximum value of $\tilde \theta$), and thus
\begin{equation} \label{s4}
\begin{split}
        \alpha_{c} = \frac{1}{\pi f |\log f|}.
\end{split}
\end{equation}
This storage capacity coincide with the optimal capacity obtained by Gutfreund for the Ising interaction case (see Appendix IV for details). 

We next ask the question of how close the threshold can be to 1, when the coding level $f$ is small but finite. The threshold needs to be sufficiently far
from one, so that the argument of the function $\Phi$ in the first condition in Eq.~(\ref{approx1}) is large and negative. We find that
for thresholds that are close to 1, the maximal capacity is
\begin{equation} \label{asy_l5}
\begin{split}
        \alpha = \frac{(1- \tilde \theta)^2}{ \pi f \left|\log (1- \tilde \theta)\right|)}.
\end{split}
\end{equation}
Eqs.(\ref{s3},\ref{asy_l5}) give the optimal threshold $\theta_{opt}$ (i.e., the optimal value of $\tilde \theta$) as a solution to the equation
\begin{equation} \label{asy_l6}
\begin{split}
        \frac{2\theta_{opt}^2 \left|\log(1-\theta_{opt})\right|}{(1-\theta_{opt})^2} = \left|\log f \right|,
\end{split}
\end{equation}
The maximum storage capacity  is
\begin{equation} \label{opt_alpha}
\begin{split}
        \alpha_c \simeq \frac{\theta^2_{opt}}{\pi f |\log f|}.
\end{split}
\end{equation}
where $\theta_{opt}$ is given as a function of $f$ by Eq.~(\ref{asy_l6}).
When $f \to 0$, $\theta_{opt} \to 1$ and Eq.(\ref{opt_alpha}) becomes the Gutfreund bound Eq.(\ref{s4}). However, this convergence is extremely slow,
as shown in Fig.~\ref{TM2}.

\subsection*{IV. Bounds for capacity}
\textbf{Gardner Upper Bound for networks with Continuous weights (GUBC).} This bound was calculated by Gardner in 1987 for networks with continuous weights. The upper bound is obtained when the volume of the space of solutions for the weights $\{J_{i j}\}$ vanishes (see details in \cite{Gardner}). By solving the Eq.(37) and Eq.(38) in \cite{Gardner}, one can obtain the GUBC for arbitrary coding levels. This result is shown by the red curve in Fig.\ref{TM2}.

In the sparse coding limit, the asymptotic solution is given in Eq.(40) in \cite{Gardner}, where:
\begin{equation} \label{u1}
\begin{split}
        \alpha_{cmax} = \frac{1}{2f|\mbox{ln} f|}.
\end{split}
\end{equation}

\noindent \textbf{Gutfreund and Stein Upper Bound for networks with Binary weights (GUBB).} This bound was calculated by Gutfreund and Stein in 1990 \cite{Gutfreund}. They extended Gardner's formalism to the case of networks with binary weights. Using a replica symmetric ansatz, the solution space of binary weights vanishes when capacity reaches:
\begin{equation} \label{u2}
\begin{split}
        \alpha_{cmax} = \frac{2}{\pi} GUBC
\end{split}
\end{equation}
In the sparse coding limit, Eq.(\ref{u1}) and Eq.(\ref{u2}) give:
\begin{equation} \label{u3}
\begin{split}
        \alpha_{cmax} = \frac{1}{\pi f|\mbox{ln} f|}.
\end{split}
\end{equation}
However, it can be shown that replica symmetry is broken, and Eq.(\ref{u2}) is an overestimate. A better estimate of the upper bound for networks with binary weights is given by zero entropy condition (see \cite{Gutfreund} for details), obtained by solving Eqs.(20-24) and Eqs.(29-34) in \cite{Gutfreund}. This zero entropy line is shown by the blue curve in Fig.\ref{TM2}. 

By numerically solving Eqs.(20-24, 29-34) in \cite{Gutfreund} and Eqs.(37-38) in \cite{Gardner}, one can see that the zero entropy line is getting close to the Gardner line (Eq.(\ref{u2})) when the coding level decrease. This numerical result indicates that Eq.(\ref{u3}) is also a good estimate for the zero entropy line in the sparse coding limit.

\subsection*{V. Storage capacity of the sparsely-connected CTF model}

Using similar calculations as in Appendix I, one can obtain that the nonlinear learning rule Eq.(\ref{M:slr}) can be transformed to a linear learning rule:
\begin{equation} \label{lslr}
\begin{aligned}
    W_{i j} = \frac{c_{ij}}{cNf(1-f)} \sum_{\mu}^{p}\left (\eta_i^{\mu} - f\right ) \left (\eta_j^{\mu} - f\right ) + \delta_{ij},
\end{aligned}
\end{equation}
where the variance of $\delta_{ij} = \frac{\alpha \Delta_0^2}{N} = \frac{\alpha}{N}(1+\frac{\pi}{2})$, and where the connection probability $c \ll 1$. In this case, the local field $h_i$ can be written as:

\begin{equation} \label{lf}
\begin{aligned}
    h_i &= \sum_{i \neq j}^N W_{ij}V_j\\
        &= \frac{1}{cf(1-f)N}\tilde{\eta}_i^1\sum_j^N c_{ij} \tilde{\eta}_j^1 V_j\\
        &+ \frac{1}{cf(1-f)N} \sum_{\mu>1}\sum_j^N c_{ij} \tilde{\eta}_i^{\mu} \tilde{\eta}_j^{\mu} V_j\\
        &+ \sum_j^N \delta_{i j} V_j,
\end{aligned}
\end{equation}
where $\tilde{\eta}_i^1$ denote the pattern that is currently being retrieved by the network. When $N$ and $p$ are large, the second and third term in the r.h.s.~of Eq.~(\ref{lf}) $Y_1$ and follow a Gaussian distribution with zero mean. Introducing order parameters $\tilde{m} = m/f(1-f)$ and $q$ defined in Eq.(\ref{M:smf}), Eq.(\ref{lf}) can be simplified to:  
\begin{equation}  \label{slf}
\begin{aligned}
    h_i &= \tilde{\eta}_i^1 \tilde{m} + Y,
\end{aligned}
\end{equation}
where the variance of $Y$ is:
\begin{equation}
\begin{aligned}
    var(Y) &= \alpha_s (\Delta_0^2+1) q
\end{aligned}
\end{equation}
and $\alpha_s = p/cN$. 
\begin{equation}  \label{scm}
\begin{aligned}
    \tilde{m} &= \frac{1}{N f(1-f)} \sum_j^N \tilde{\eta}_j^1 \Theta(\tilde{\eta}_j^1 \tilde{m} + Y - \theta),
\end{aligned}
\end{equation}

\begin{equation}  \label{scq}
\begin{aligned}
    q  &= \frac{1}{N} \sum_j^N  \Theta^2(\tilde{\eta}_j^1 \tilde{m} + Y - \theta).
\end{aligned}
\end{equation}
Averaging Eq.(\ref{scm},\ref{scq}) over $\eta$ and the Gaussian noise $Y$, the mean-field equations for order parameters $\tilde{m}$ and $q$ are
\begin{equation} \label{smf}
\begin{split}
       &\tilde{m} = \Phi \left(  \frac{\tilde \theta - \tilde{m}(1-f)}{\sqrt{\alpha_s q(1+ \Delta_0^2)}}\right) - \Phi \left(  \frac{ \tilde \theta + \tilde{m}f}{\sqrt{\alpha_s q(1+ \Delta_0^2)}}\right),\\
       &q = f\Phi \left(  \frac{ \tilde \theta - \tilde{m} (1-f)}{\sqrt{\alpha_s q (1+ \Delta_0^2)}}\right) + (1-f)\Phi \left(  \frac{\tilde \theta + \tilde{m}f}{\sqrt{\alpha_s q(1+ \Delta_0^2)}}\right),
\end{split}
\end{equation}
Note that these equations coincide with the fully connected case in the sparse coding limit, as in the case of continuous weights \cite{Tsodyks,Tsodyks_2}.

\subsection*{VI. Numerical simulations}
The simulations in Fig.\ref{TM1} used a network with 4,000 neurons and coding level $f=0.02$. The overlaps $\tilde{m}$ are averaged over five independent realizations. Simulations consist in a learning phase in which the connectivity matrix $W_{ij}$ is built, and a retrieval phase in which network dynamics run until it reaches a fixed point. 
For each input pattern $\mu$, we choose as initial conditions $\{V_i=\eta_i^{\mu}\}$. 
Overlaps $m$ are obtained by averaging over all $m^{\mu}$.


\begin{thebibliography}{99}
\bibitem{Hopfield} Hopfield, John J. "Neural networks and physical systems with emergent collective computational abilities." Proceedings of the national academy of sciences 79.8  (1982): 2554-2558.
\bibitem{Tsodyks} Tsodyks, Mikhail V., and Mikhail V. Feigel'man. "The enhanced storage capacity in neural networks with low activity level." EPL  (Europhysics Letters ) 6.2  (1988): 101.
\bibitem{Amit} Amit, Daniel J., Hanoch Gutfreund, and Haim Sompolinsky. "Statistical mechanics of neural networks near saturation." Annals of physics 173.1  (1987): 30-67.
\bibitem{Gardner} Gardner, Elizabeth. "The space of interactions in neural network models." Journal of physics A: Mathematical and general 21.1  (1988): 257.
\bibitem{Petersen} Petersen, Carl CH, et al. "All-or-none potentiation at CA3-CA1 synapses." Proceedings of the National Academy of Sciences 95.8  (1998): 4732-4737.
\bibitem{Connor} O'Connor, Daniel H., Gayle M. Wittenberg, and Samuel S-H. Wang. "Graded bidirectional synaptic plasticity is composed of switch-like unitary events." Proceedings of the National Academy of Sciences 102.27  (2005): 9679-9684.
\bibitem{Hruska} Hruska, Martin, et al. "Synaptic nanomodules underlie the organization and plasticity of spine synapses." Nature neuroscience 21.5  (2018): 671-682.
\bibitem{Dork} Dorkenwald, Sven, et al. "Binary and analog variation of synapses between cortical pyramidal neurons." BioRxiv  (2019).
\bibitem{Krauth} Krauth, Werner, and Marc M\'{e}zard. "Storage capacity of memory networks with binary couplings." Journal de Physique 50.20 (1989): 3057-3066.
\bibitem{Sompolinsky_1} Sompolinsky, Haim. "The theory of neural networks: The Hebb rule and beyond." Heidelberg colloquium on glassy dynamics. Springer, Berlin, Heidelberg, 1987.
\bibitem{Sompolinsky_2} Sompolinsky, Haim. "Neural networks with nonlinear synapses and a static noise." Physical Review A 34.3 (1986): 2571.
\bibitem{Lee} Lee, Jae Sung, et al. "The statistical structure of the hippocampal code for space as a function of time, context, and value." Cell 183.3 (2020): 620-635.
\bibitem{Waydo} Waydo, Stephen, et al. "Sparse representation in the human medial temporal lobe." Journal of Neuroscience 26.40 (2006): 10232-10234.
\bibitem{Gutfreund} Gutfreund, Hanoch, and Yaakov Stein. "Capacity of neural networks with discrete synaptic couplings." Journal of Physics A: Mathematical and General 23.12  (1990): 2613.
\bibitem{Mezard} M\'{e}zard, M., Parisi, G., and Virasoro, M. A. (1987). Spin glass theory and beyond: An Introduction to the Replica Method and Its Applications (Vol. 9). World Scientific Publishing Company.
\bibitem{Markram}  Markram, Henry, et al. "Physiology and anatomy of synaptic connections between thick tufted pyramidal neurones in the developing rat neocortex." The Journal of physiology 500.2 (1997): 409-440.
\bibitem{Tsodyks_2} Tsodyks, M. V. "Associative memory in asymmetric diluted network with low level of activity." EPL (Europhysics Letters) 7.3 (1988): 203.
\bibitem{Derrida} Derrida, Bernard, Elizabeth Gardner, and Anne Zippelius. "An exactly solvable asymmetric neural network model." EPL (Europhysics Letters) 4.2 (1987): 167.

\bibitem{Tsodyks_3} Tsodyks, M. V. "Associative memory in neural networks with binary synapses"  Modern Physics Letters B 4 (11), 713-71613..
\bibitem{Amit_Fusi_1}  Amit, Daniel J., and Stefano Fusi. "Constraints on learning in dynamic synapses." Network: Computation in Neural Systems 3.4 (1992): 443-464.
\bibitem{Amit_Fusi_2} Amit, Daniel J., and Stefano Fusi. "Learning in neural networks with material synapses." Neural Computation 6.5 (1994): 957-982.
\bibitem{Dubreuil}  Dubreuil, Alexis M., Yali Amit, and Nicolas Brunel. "Memory capacity of networks with stochastic binary synapses." PLoS computational biology 10.8 (2014): e1003727.
\bibitem{Benna_Fusi} Benna, Marcus K., and Stefano Fusi. "Computational principles of synaptic memory consolidation." Nature neuroscience 19.12 (2016): 1697-1706.
\bibitem{Ganguli} Lahiri, Subhaneil, and Surya Ganguli. "A memory frontier for complex synapses." Advances in neural information processing systems 26 (2013): 1034-1042.


\bibitem{Fusi2005} Fusi, Stefano, Patrick J. Drew, and Larry F. Abbott. "Cascade models of synaptically stored memories." Neuron 45.4 (2005): 599-611.

\bibitem{Fusi2007}  Fusi S, Abbott LF. Limits on the memory storage capacity of bounded synapses.
Nat Neurosci. 2007 Apr;10(4):485-93
\end{thebibliography}
\end{document}